\documentclass[tikz,fleqn,usenatbib]{aa}

\usepackage{lastpage}
\usepackage{graphicx}
\usepackage{dcolumn}
\usepackage{bm}
\usepackage{natbib}
\usepackage{multirow}
\usepackage{url}
\usepackage{color}
\usepackage{xcolor}
\usepackage{float}
\usepackage{cancel}
\usepackage{amssymb}
\usepackage{amsmath}
\usepackage{colortbl} 
\usepackage{comment}
\usepackage{mathptmx}
\usepackage{txfonts}
\usepackage[T1]{fontenc}
\usepackage{ae,aecompl}
\usepackage{orcidlink}
\usepackage{academicons}
\usepackage{hyperref}
\usepackage{cleveref}
\usepackage{adjustbox}
\usepackage{amssymb}
\usepackage{latexsym}
\usepackage{natbib,twoopt}
\usepackage[normalem]{ulem}

\newcommand{\be}{\begin{equation}}
\newcommand{\vr}{\textbf{r}}
\newcommand{\ee}{\end{equation}}
\newcommand{\vk}{\textbf{k}}

\begin{document}

\title{Individual halo bias in models of $f(R)$ gravity}
\titlerunning{J.E García-Farieta et al.}
\author{
J.E Garc\'ia-Farieta$^{1}$\orcidlink{0000-0001-6667-5471}, Antonio D. Montero-Dorta$^{2}$\orcidlink{0000-0003-4056-2246} and Andrés Balaguera-Antolínez$^{3,4}$\orcidlink{0000-0001-5028-3035}}
\institute{
 Departamento de F\'isica, Universidad de C\'ordoba, E-14071, Córdoba, Spain \and
 Departamento de F\'isica, Universidad T\'ecnica Federico Santa Mar\'ia, Avenida Vicu\~na Mackenna 3939, San Joaqu\'in, Santiago, Chile \and 
 Instituto de Astrof\'{\i}sica de Canarias, s/n, E-38205, La Laguna, Tenerife, Spain \and 
 Departamento de Astrof\'{\i}sica, Universidad de La Laguna, E-38206, La Laguna, Tenerife, Spain
  }
\authorrunning{Author}

\date{Received /Accepted}

\abstract
{Halo bias links the statistical properties of the spatial distribution of dark matter halos to those of the underlying dark matter field, providing insights into clustering properties in both general relativity (GR) and modified-gravity scenarios such as $f(R)$ models. While the primary halo mass-dependent bias has been studied in detailed, the secondary bias, which accounts for the additional dependencies on other internal halo properties, can offer a sensitive probe for testing gravity beyond the $\Lambda$CDM model.}
{To quantify any potential deviations between $\Lambda$CDM and $f(R)$ gravity models in halo clustering, at both the primary and secondary level, as well as in the distributions of halo properties in the cosmic web.}
{Using $N$-body simulations of $f(R)$ gravity models, we assess the scaling relations and the primary and secondary bias signals of halo populations on the basis of a halo-by-halo estimator of large-scale effective bias. Our analysis is performed using halo number density as the independent variable.}
{The relative difference in the effective bias between the $f(R)$ models and $\Lambda$CDM is sensitive, albeit slightly, to the power index of modified gravity. The largest deviations from GR are measured for low-mass halos, where the average bias at fixed number density decreases by up to 5\% for fixed scaling indices. We also show that the scaling relations for some environmental properties, including neighbour statistics, Mach number and local overdensity, exhibit small but non-negligible deviations (~3-5\%) from GR for a wide range of number densities. Our results also suggests that the properties of halos in sheets and voids show the largest departures from GR (> 10\% in some cases). In terms of secondary bias, we do not find any statistically significant deviations with respect to $\Lambda$CDM for any of the properties explored in this work.}
{}
\keywords{
Cosmology: large-scale structure of Universe \--- Methods: numerical \--- Galaxies: halos}
\maketitle

\section{Introduction}\label{sec:intro}

The physical origin of the late-time accelerated expansion of the Universe remains one of the most pressing open problems in modern cosmology. This challenge is further compounded by the latest cosmological constraints from the Dark Energy Spectroscopic Instrument (DESI) \citep{2025arXiv250314743D}. Cosmic acceleration can potentially be explained through two primary avenues, either by invoking dark energy models characterised by an equation of state that generates a negative pressure, possibly with a time-dependent variation, or alternatively by modifying General Relativity (GR), a framework that remains largely untested on cosmological scales.

Recent studies show that while modified gravity (MG) does not unequivocally outperform the standard $\Lambda$CDM ($\Lambda$-cold dark matter) model, it offers a robust physical mechanism for safely crossing the phantom divide. This is investigated in the context of the general scalar-tensor Horndeski theory, a MG framework that encompasses the widely studied $f(R)$ theory \citep[see, e.g.,][]{2024PhRvD.110l3524C}. The $f(R)$ models, extensively explored in the literature \citep[for a concise review see][]{2016RPPh...79d6902K, Nojiri2017}, are of particular interest because they represent a minimal yet significant modification of GR. A family of $f(R)$ models, including the popular \citet{2007PhRvD..76f4004H} parametrization, in light of DESI data, has also been tested alongside other cosmological data \citep{2024arXiv241112026I, 2025arXiv250316132P}. At the same time, forecasts from the Euclid space mission \citep{2011arXiv1110.3193L} suggest that it will be capable of distinguishing $f(R)$ gravity from the $\Lambda$CDM model with a confidence level exceeding $3\sigma$ by combining spectroscopic and photometric probes \citep[][]{2023arXiv230611053C, 2024A&A...690A.133F}. Additional insights into $f(R)$ gravity, derived from the full-shape galaxy power spectrum, are discussed in \cite{2025PhRvD.111b1301A}.

Beyond its potential to explain cosmic acceleration and provide an accurate description of cosmic microwave background (CMB) observations and galaxy clustering \citep{Berti2015}, $f(R)$ gravity presents a compelling alternative to dark energy models by preserving the source tensor while introducing modifications to the field equations themselves. Plausible modifications of gravity employ a screening mechanism that suppresses the effect of the fifth force in regimes where strong constraints on gravity are set by observations, such as the solar system. In this study, we consider the \citet{2007PhRvD..76f4004H} parametrization of $f(R)$, combined with a screening mechanism as described by \citet{2004PhRvD..69d4026K}. In this framework, the modification of gravity manifests as the addition of a scalar field coupled to matter, which introduces changes in the density field. As a result, deviations from GR are expected to appear in the clustering properties of biased cosmic tracers, such as galaxies, dark matter halos, and quasars. A variety of approaches have been employed with the aim of distinguishing potential variations in gravity from the standard $\Lambda$CDM model, with efforts including the study of redshift-space distortions \citep{Jennings2012, 2019JCAP...06..040W, GF2019, HernandezAguayo2019, GF2020, 2022JCAP...12..028F,2024arXiv241205662L}, tracer bias and sample selection \citep{2017MNRAS.467.1569A, GF2021}, bispectrum of lensing \citep{2025arXiv250309893C} and matter \citep{2011JCAP...11..019G}, cosmic voids \citep{2015MNRAS.451.4215Z, Voivodic2017, Perico2017, Contarini2021}, halo mass functions \citep{Hagstotz2019, Gupta2022} and a variety of complementary methodologies.

In this work, we investigate the effect of assuming different $f(R)$ models on the linear bias of dark-matter (DM) halos, focusing on its multiple dependencies. The term secondary halo bias is used to represent all dependencies of halo bias that are measured at fixed halo mass; the dependence of bias on halo (virial) mass is considered here the primary dependence, arising from the fundamental connection between bias and the peak height of density fluctuations (see, e.g.,\citealt{Kaiser1984,Bardeen1986,Mo1996, ShethTormen1999, Sheth2001, Tinker2010}). Secondary bias have been measured for a large number of internal halo properties, including concentration, formation time, shape or spin, to name but a few\footnote{The secondary dependencies directly related to the different accretion histories of halos are specifically called {\it{halo assembly bias}}.} (see, e.g., \citealt{Sheth2004,Gao2005,Wechsler2006,Gao2007,Dalal2008, Angulo2008,Li2008,Faltenbacher2010, Lazeyras2017,2018Salcedo,Mao2018, Han2018,SatoPolito2019, Johnson2019, Ramakrishnan2019,Contreras2019, MonteroDorta2020B, Tucci2021,  Contreras2021_cosmo, MonteroDorta2021, Balaguera2024, MonteroRodriguez2024, MonteroDorta2025}). 

At fixed halo mass, the bias also correlates with the properties of the environment that halos inhabit, including the local overdensity, the anisotropy of the tidal field, the distance to cosmic-web features or even the Mach number (e.g., \citealt{2018MNRAS.476.3631P, MonteroRodriguez2024, Balaguera2024}). The significance of some these environmental properties resides in the fact that they can be seen as  
mediators linking bias and the internal properties of halos \citep{2018MNRAS.476.3631P, Ramakrishnan2019}. In \cite{Balaguera2024}, we analysed the secondary bias signals produced by a variety of internal and environmental properties, along with the dependence of the trends on the location of halos within the cosmic web. 

To facilitate the analysis of bias for different $f(R)$ models, we use a halo-by-halo bias estimator \citep{2018MNRAS.476.3631P, Contreras2021a,Balaguera2024, MonteroDorta2025}, which provides significant analytical advantages as compared to the traditional approach based on ratios of correlation functions or power spectra (see \citealt{Balaguera2024} for more information). 

The outline of this paper is as follows. In Sec.~\ref{sec:simu}, we briefly describe the background theory, summarizing the modified gravity model we consider and describing our set of simulations. Section~\ref{sec:method} details the method used to assign the individual effective halo bias and its application to the $f(R)$ simulations. The results of the effective bias, as well as the primary and secondary bias relations measured in the $f(R)$ halo catalogs, are presented in Sec.~\ref{sec:results}. Finally, in Sec.~\ref{sec:conclusions}, we summarize our main findings and discuss the implications of our results.

\section{Models and simulations}\label{sec:simu}

The $f(R)$ theory of gravity is a notable example of a non-standard cosmological model developed to address some of the issues present in the $\Lambda$CDM model \citep[we refer the reader to][for in-depth details]{2012PhR...513....1C,2010RvMP...82..451S}. Among the various parametrizations of this theory, the \citet{2007PhRvD..76f4004H} one is particularly well-studied due to its physical motivation and the fact that it remains a viable model, not yet ruled out by observations \citep{2025PhRvD.111b1301A,2025arXiv250111772D,2016PhRvL.117e1101L,2024MNRAS.534..349L,2024A&A...691A.301A,2025PhRvD.111d3519V}. This model modifies the Einstein-Hilbert action by replacing the Ricci scalar with a function $f(R)$, such that $R\mapsto R + f(R)$. The function $f(R)$ can be adjusted to match, to first order, the background expansion of the $\Lambda$CDM model and, for the Hu-Sawicki parametrization, it is modelled as a broken power law of the Ricci scalar, given by
\begin{equation}
f(R)=-2 \Lambda \frac{R^n}{R^n+\mu^{2 n}},
\end{equation}
where $\Lambda$ and $\mu^2$ serve as free parameters for each value of the power index $n$. In the limit of small deviations from GR, i.e., $R \gg \mu$, the $f(R)$ function is approximately equal to
\begin{equation}
f(R)=-2 \Lambda-\frac{\lvert f_{R 0} \rvert}{n} \frac{R_0^{n+1}}{R^n},
\end{equation}
where $R_0$ denotes the background Ricci scalar at the current epoch and $f_{R 0}$ is the derivative of $f(R)$ with respect to $R$ (i.e., $f_R \equiv \frac{d}{dR}f(R)$) evaluated at the present time. This dynamical degree of freedom is typically interpreted as a scalar field and can be approximated as $f_{R0} = -2n \Lambda \mu^{2n} / R_0^{n+1}$. Note that if $|f_{R0}| \to 0$, the boundary condition $f(R) \to -2\Lambda$ is satisfied as a consequence of requiring equivalence with $\Lambda$CDM. We recall that here $n \geq 0$ is an integer, and while most past simulations in $f(R)$ have focused on the scenario where $n = 1$, the scenario with $n = 2$ has not been as thoroughly investigated. This paper examines both scenarios within its set of simulations, comparing the results against the standard $\Lambda$CDM model.

We use the MG simulation suite described in \citet[][]{2024A&A...690A..27G}, which consists of a set of seven high-resolution COLA simulations performed with the \texttt{FML-COLA} solver\footnote{\url{https://github.com/HAWinther/FML/tree/master/FML/COLASolver}}. The simulations are consistent with the best-fit parameters of Planck 2018 cosmology: $\Omega_\text{m} = 0.311$, $\Omega_{\text{cdm}} = 0.2621$, $\Omega_\text{b} = 0.0489$, $h = 0.6766$, $n_s = 0.9665$, and $A_s = 2.105 \times 10^{-9}$ \citep{2020A&A...641A...6P}. They feature the dynamics of $2048^3$ dark matter particles in a comoving box of $1 \, h^{-1}\text{Gpc}$ on each side. The simulations are started at redshift $z = 99$, at which time non-linear effects are negligible, and evolved up to redshift $z = 0.5$ using 100 time steps. All the MG models use the same initial conditions as $\Lambda$CDM, which were generated using the \texttt{2LPTic} code \citep{1998MNRAS.299.1097S,2006MNRAS.373..369C} implemented in \texttt{FML-COLA}. This setup achieves a mass resolution of $M_p = 1.005 \times 10^{10}~h^{-1}\mathrm{M}_{\odot}$, ensuring sufficient detail for accurate halo identification. We use the \texttt{ROCKSTAR} halo-finding algorithm \citep{2013ApJ...762..109B} to identify halos in the simulation containing at least 80 dark matter particles each. To mitigate the small-scale uncertainty of COLA, we adjusted the force resolution in the \texttt{ROCKSTAR} parameter file to be close to the grid size of the simulations. To focus on more massive structures, we apply an additional mass threshold of $10^{12}$~$h^{-1}\mathrm{M}_{\odot}$, retaining only halos exceeding this limit. For our analysis, we considered only the MG models that are not completely ruled out by current constraints, leaving us with four models: $(|f_{R0}|,\,n) \in \{(10^{-5},\,1),\,(10^{-6},\,1),\,(10^{-5},\,2),\,(10^{-6.5},\,2)\}$, which are denoted as $\{F5_1,~F6_1,~F5_2,~F6.5_2\}$.

\begin{figure*}
   \centering
   \includegraphics[width=\linewidth]{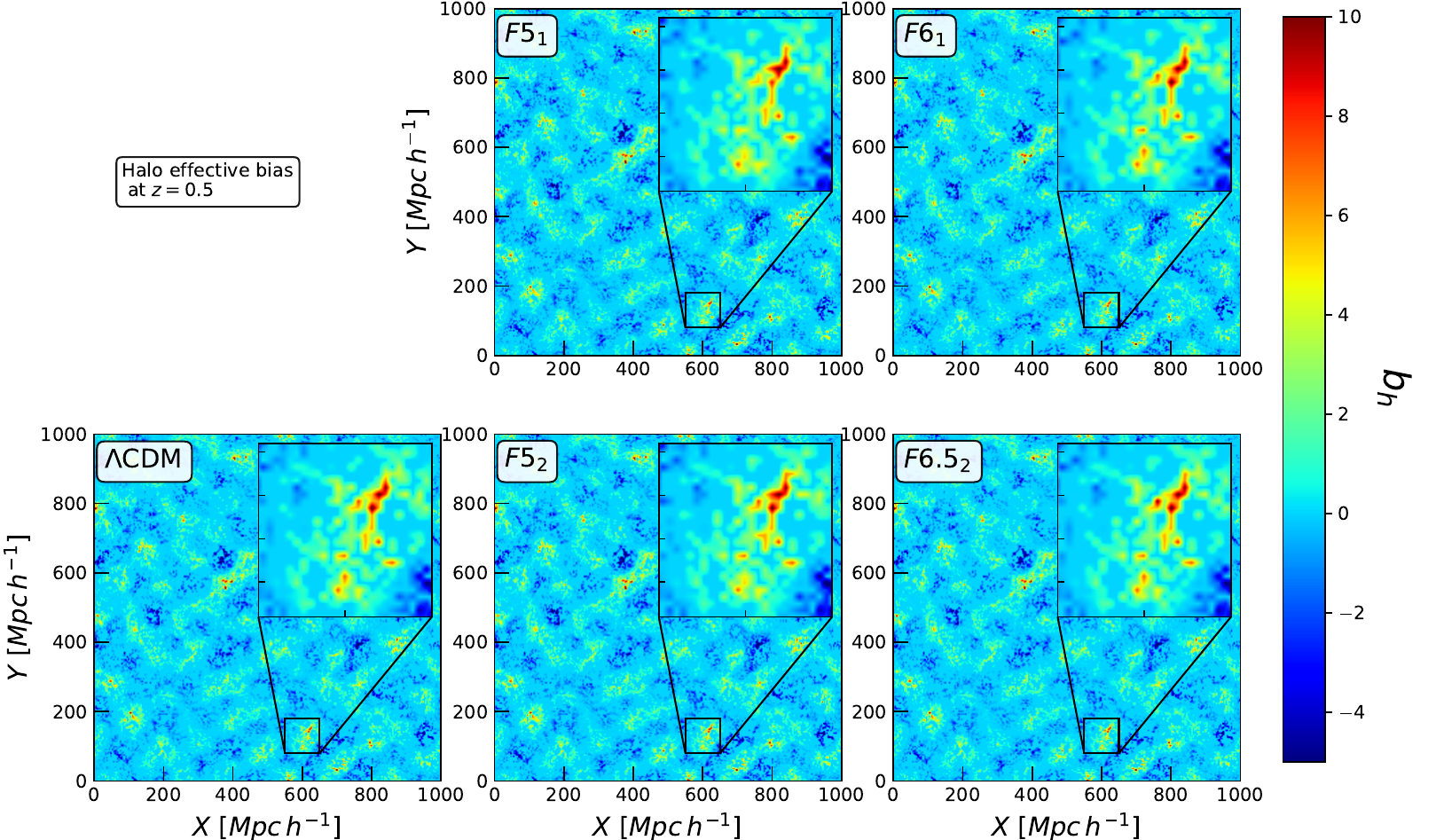}
      \caption{
      \small{Slices of the $\sim12\,h^{-1}$Mpc thickness in the MG simulations (at $z=0.5$) showing the individual bias field. The zoom-in region represents a $100\times100$\,$h^{-2}$Mpc$^{-2}$ slice}. 
      }
      \label{fig:figure1}
\end{figure*}

\section{Assignment of individual effective bias}\label{sec:method}
We implement the approach of \citet[][]{2018MNRAS.476.3631P}, in which large-scale (or effective) bias can be assigned to individual tracers in a simulation. This approach implements the estimator for large-scale bias based on the cross-correlation between the halo tracer field and its underlying dark matter density field \citep[see e.g.][]{2007PhRvD..75f3512S, 2012MNRAS.420.3469P,2014A&A...563A.141B}. Briefly, for a dark matter tracer at position $\vr_{i}$, we assign individual bias as \footnote{The assignment of individual halo bias has been performed using the \texttt{CosmiCCcodes} library at \url{https://github.com/balaguera/CosmicCodes}.}
\be\label{eq:bias_object}
b^{(i)}_{\mathrm{eff}}=\frac{\sum_{j,k_{j}<k_{max}}N^{j}_{k}\langle {\rm e}^{-i\vk \cdot \vr_{i}} \delta_{\mathrm{dm}}^{*}(\vk) \rangle_{k_{j}}}{\sum_{j,k_{j}<k_{max}} N^{j}_{k}P_{\rm dm}(k_{j})},
\ee
where $\delta_{\mathrm{dm}}(\vk)$ is the Fourier transform of the underlying dark matter density field, $P_{\rm dm}(k_{j})$ its power spectrum and $N^{j}_{k}$ is the number of Fourier modes in the $j$-the spherical bin in Fourier space \citep[see e.g.][]{Ramakrishnan2019,2019MNRAS.482.1900H,2020MNRAS.495.3233P,2021MNRAS.504.5205C,Balaguera2024,Balaguera2024_2, MonteroDorta2025}. The sum is carried over the range of wavenumbers $k_{j}<k_{max}$ in which the ratio between the halo and the dark matter power spectra is constant\footnote{We have used a maximum wavenumber $k_\mathrm{max}=0.08$ $h$Mpc$^{-1}$, up to which the ratio $P_{h}(k)/P_\mathrm{dm}(k)$ is compatible with a constant value at the current redshift.}. \citet[][]{Balaguera2024} showed that the effective bias as a function of multiple halo properties, as obtained from standard approaches (for example, measurements of the auto and cross-power halo power spectrum in bins of a that halo property, see e.g., \citet[][]{2014A&A...563A.141B,2014MNRAS.440..555P}), is consistent with the results obtained from Eq.~(\ref{eq:bias_object}) and with known calibrations in the literature, such as the halo bias - mass relation (\citealt{2010ApJ...724..878T}, see also \citealt{2018MNRAS.476.3631P}). This way of assigning bias to objects has also been proven advantageous as a means of including the clustering signal (up to a given scale $k_{\max}$) in the machinery for assigning halo properties in the context of so-called calibrated methods, as discussed by \citet[][]{Balaguera2024_2}. 

Figure~\ref{fig:figure1} displays maps of the halo effective bias in slices of 12~$h^{-1} \text{Mpc}$ thickness, which vary spatially under different MG models, as labelled. The individual halo effective bias is computed using equation Eq.~(\ref{eq:bias_object}) for all the MG models at $z = 0.5$. A zoomed-in region, representing a portion of $100 \times 100$ $h^{-2} \text{Mpc}^2$, highlights subtle differences that are visible to the naked eye. In particular, MG models tend to alter the clustering of halos compared to the standard $\Lambda$CDM model, resulting in sharper or more fragmented structures in some cases. These differences arise from changes in the gravitational dynamics, which influence the formation and distribution of halos.

\begin{figure}
  \centering
  \includegraphics[width=\linewidth]{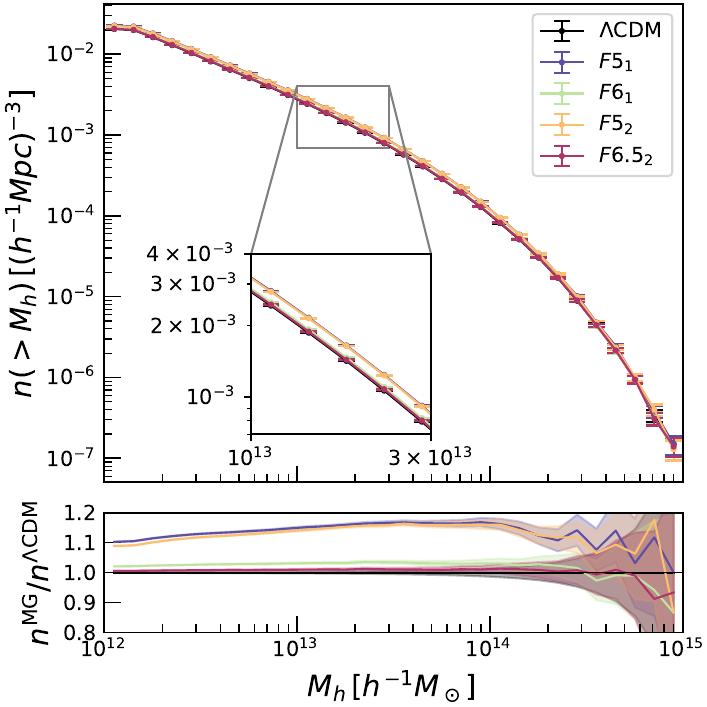}
     \caption{
     \small{Cumulative halo mass function for the different MG models taken at redshift $z=0.5$. The bottom panel shows the relative deviation with respect to the $\Lambda$CDM model with lines connecting the datapoints to guide the eye.}
     }
    \label{fig:figure0}
\end{figure}

\section{Results}\label{sec:results}

In this analysis, and to facilitate a fair comparison among the different MG models, we chose to use number density as the independent variable (i.e., the primary halo property) rather than halo mass, as the latter is model-dependent. This choice is motivated by previous research, such as \cite{2017MNRAS.467.1569A} and \cite{GF2021}, where the clustering statistics of MG models are analysed in terms of halo populations defined by fixing the number density for each catalog. This approach is essentially a simplified version of halo abundance matching. Figure~\ref{fig:figure0} shows the cumulative halo mass function (HMF) $\bar{n}(>M)$ for the different $f(R)$ models. As demonstrated by the HMF trend, $f(R)$ models predict a higher number of halos than the $\Lambda$CDM model for masses above $10^{12}~h^{-1}\mathrm{M}_{\odot}$ due to the enhanced gravitational effects. The zoom-in section of Fig.~\ref{fig:figure0} highlights the intermediate mass range between $10^{13}$ and $3 \times 10^{13}$ $h^{-1}\mathrm{M}_{\odot}$, where all models show slightly different HMF values. As expected, the $F6$ family of models closely aligns with $\Lambda$CDM, as they exhibit the least gravity enhancement, resulting in a halo mass distribution that resembles that of $\Lambda$CDM, with deviations of less than 5\%, as shown in the lower panel of Fig.~\ref{fig:figure0}. In contrast, the remaining $f(R)$ models, specifically the $F5$ models, show an excess in halo mass, producing 10 to 15\% more halos than predicted by $\Lambda$CDM. As a consequence, the mass cuts corresponding to a given number density vary across the various $f(R)$ catalogs. Therefore, from this point onward, we will present our results in terms of number density, $\bar{n}$. This choice is made for the sake of a fair comparison among models and is also motivated by the fact that the virial halo mass is not an observable. Therefore, a sample selection with a fixed $M_{\text{min}}$ cannot be directly replicated in a real galaxy sample.

\begin{figure}
  \centering
  \includegraphics[width=\linewidth]{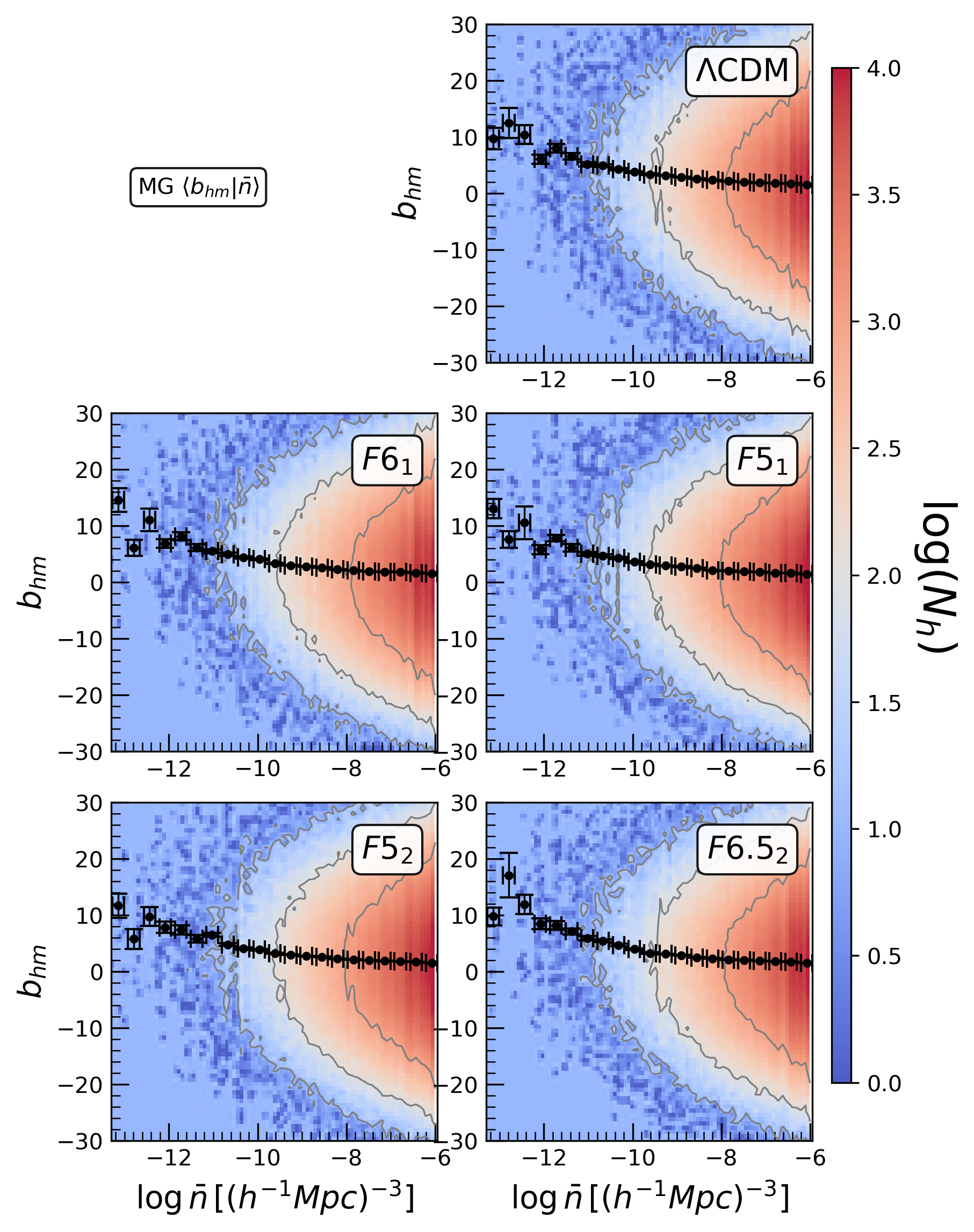}
     \caption{
     \small{Halo effective bias $b_{hm}$ computed at $z=0.5$ as a function of the number density. The points denote the mean bias in different mass bins with errorbars denoting the standard error of the mean. The contours indicate a region of an equal number of tracers $N_h$.}
     }
    \label{fig:figure2}
\end{figure}

\subsection{Primary bias: scaling relations}
We obtain the large-scale halo bias using the object-by-object estimator of Eq.~\eqref{eq:bias_object}, which offers an approach to investigate its dependencies on several halo properties as presented in \cite{Balaguera2024}. Figure~\ref{fig:figure2} shows the individual halo effective bias as a function of number density for different $f(R)$ models at $z=0.5$. The black points represent the mean bias within a number density bin, with error bars computed as the standard error of the mean for each bin. The colour bar in Fig.~\ref{fig:figure2} represents the number of tracers, $N_h$, while the contours outline regions with an equal number of tracers. As we see in the figure, the effective bias for the MG models follows a trend similar to that of the baseline $\Lambda$CDM model. However, slight differences are observed at low number densities in the the $f(R)$ models, mainly due to the large scatter introduced by very massive halos.
Notably, models such as $F_{52}$ and $F_{6.52}$ exhibit more pronounced shifts and deeper dips, indicating stronger deviations from standard gravity effects at lower number densities ($\log \bar{n} < -10$).
\begin{figure}
  \centering
  \includegraphics[width=0.9\linewidth]{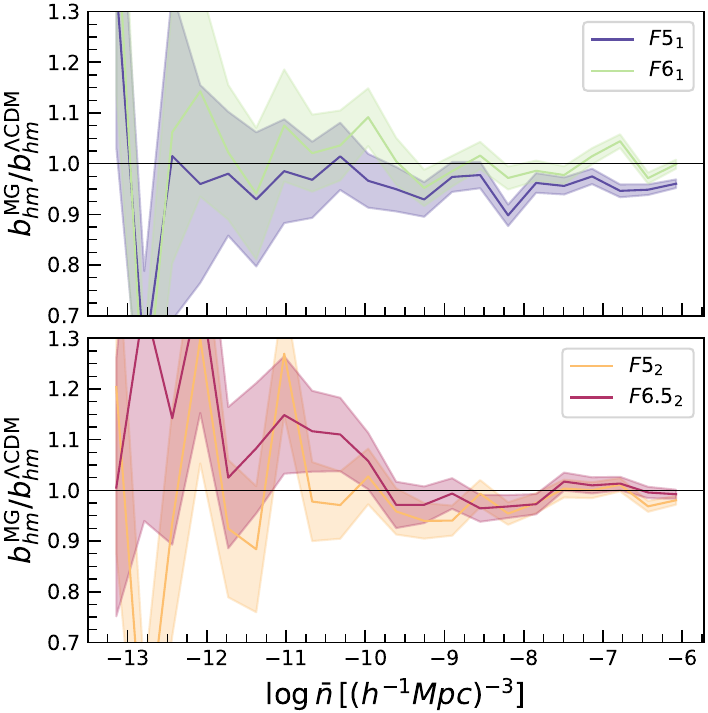}
     \caption{
     \small{Ratios of the effective halo bias in the $f(R)$ simulations at $z=0.5$, measured in bins of the halo viral mass. Symbols denote the mean halo-mass bias $\langle b_{hm}|\log M_{vir}\rangle $.}
     }
    \label{fig:figure3}
\end{figure}

In Fig.~\ref{fig:figure3}, we compare the relative deviations of the $f(R)$ effective halo bias from $\Lambda$CDM, computed as described above, with the shaded bands representing the expected error bars derived from the standard error of the mean halo bias. As the value of $|f_{R0}|$ increases, the halo bias tends to decrease because halos of a fixed number density become more common and thus less biased. For the $F5$ models, we observe the effect of the power index on the bias, with visible differences between $n=1$ and $n=2$. Specifically, for low number densities ($\log \bar{n} < -10$), no significant impact of the power index is observed, while for $\log \bar{n} > -10$, we find that $F5_1$ is more biased than $F5_2$, with a deviation from $\Lambda$CDM of about 5\% for $F5_1$ and approximately 2.5\% for $F5_2$. In contrast, both $F6$ models, with $n = 1$ and $n = 2$, are consistent with $\Lambda$CDM within the error bars across the entire number density range, with departures from $\Lambda$CDM of up to 2.5\%. The impact of the power index in these models is less clear, as they exhibit different amplitudes for the scalaron field $|f_{R0}|$.

In a study on halo statistics, \cite{2009PhRvD..79h3518S} found that in full simulations of $f(R)$ cosmologies —such as in the present work— the difference in bias is reduced compared to simulations that do not account for gravity modifications in the deep potential wells of cosmic structures. According to these findings, the largest deviations from GR for the $F5_1$ model occur in low-mass halos (high number density), while for the $F6_1$ model, the differences are more noticeable in halos of intermediate mass, rather than in the lowest mass halos. This observation is consistent with the results of our analysis.
At background field values of $|f_{R0}| \lesssim 10^{-5}$, current observations show that the enhancement of the gravitational force is suppressed inside halos, significantly reducing the effects on the properties of the most massive halos. \cite{2009PhRvD..79h3518S} has pointed out that the scaling relations still have applicability for establishing conservative upper limits on modifications to gravity, even in simpler cases where the halo bias is derived by dividing the halo cross power spectrum by the matter power spectrum. This limitation motivates our approach, which uses the object-by-object bias to set more reliable constraints on modifications to gravity.
\begin{figure*}[!htbp]
  \centering
  \includegraphics[width=\linewidth]{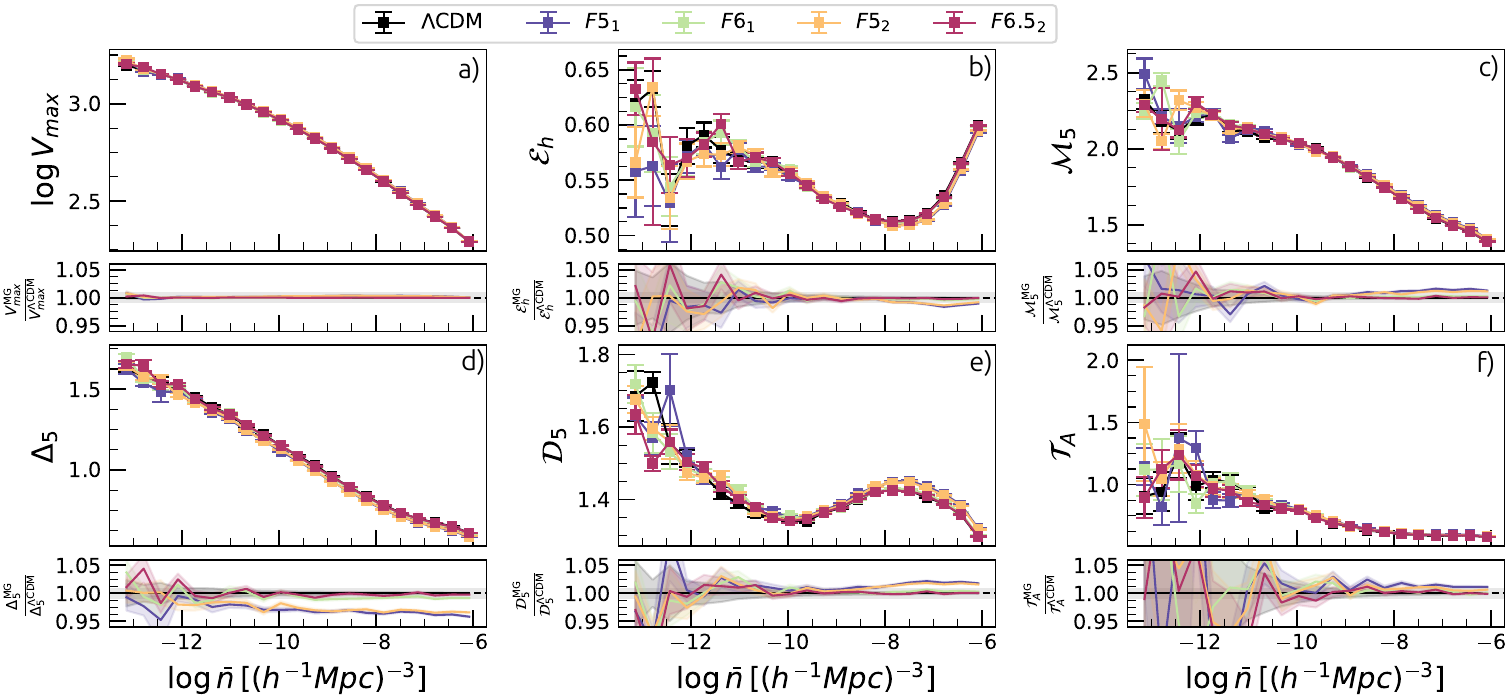}
     \caption{
     \small{Mean scaling relations between halo number density and selected non-local halo properties across different $f(R)$ cosmologies at redshift $z = 0.5$. The properties plotted are $V_{\mathrm{max}}$, $\mathcal{E}_h$, $\mathcal{M}_5$, $\Delta_5$, $\mathcal{D}_5$, and $\mathcal{T}_A$. The lower panels show the ratio of each property in $f(R)$ with respect to $\Lambda$CDM. The error bars represent the uncertainty in the mean for each mass bin.}}
    \label{fig:figure5p5}
\end{figure*}

We also investigated the behaviour of scaling relations among various halo properties across different simulations, including novel environmental properties introduced by \citet{Balaguera2024}, which are based on both the underlying dark matter field and halo properties. These scaling relations are expected to vary as halos change in mass, or equivalently, as their number density evolves. Differences in the occupancy of the bins of the scaling relations in the $f(R)$ cosmologies can arise through gravitational evolution and the time- and scale-dependence of the growth factor in these models \citep[see e.g.][]{2024A&A...690A..27G}. We assess the strength of the relationships of the halo scaling relations with secondary properties, as well as their impact on the signal of both the primary and secondary halo bias in these cosmologies. Below, we briefly describe the quantities considered in the scaling relations \citep[for more details, we refer the reader to][]{Balaguera2024}:
\begin{itemize}
    \item The \textbf{maximum circular velocity}, $V_{\text{max}}$, is defined as $V_{\text{max}} = \text{Max}(V_c)$, where $V_c(r)$ is the Newtonian circular velocity of the halo. $V_{\text{max}}$ serves as a probe of the potential well of dark matter halos. 
    \citet{2016MNRAS.462.1530A} showed, using zoomed-in simulations of Milky Way-like halos in $f(R)$ cosmologies, that the gravitational potential drops below a crucial threshold for halos with masses greater than $2.3 \times 10^{12}~h^{-1}\mathrm{M}_\odot$. This results in a rapid suppression of the fifth force of MG, leading to significant deviations from the $\Lambda$CDM model. However, at radius close $r_{200}$ (the radius that encloses a sphere with a mean density 200 times the critical density), $V_{\text{max}}$ tends to remain unchanged relative to $\Lambda$CDM due to the screening mechanism of MG.  
    \item The \textbf{halo ellipticity}, denoted as $\mathcal{E}_h$, measures the shape of the halo and is defined as the ratio of the semi-axes $a$, $b$, and $c$ of the halo’s ellipsoid, given by $\mathcal{E}_h \equiv \frac{1 - s^2}{1 + s^2 + q^2}$, where $q = b/a$ and $s = c/a$, with the condition $a > b > c$. The semi-axes $a$, $b$, and $c$ are determined using the \texttt{ROCKSTAR} halo finder. The halo ellipticity has two special cases: in the spherical limit, $\mathcal{E}_h \to 0$, and in nonspherical or aspherical configurations, $\mathcal{E}_h > 0$.
    \item The \textbf{relative local Mach number}, $\mathcal{M}_5$, is a measure of the local kinematic temperature of the tracer distribution. It quantifies the bulk motion of patches of the Universe relative to the velocity dispersion of random motions within those regions. This quantity has been shown to exhibit correlations with various galaxy and halo properties. The relative Mach number is computed from the relative velocity between a dark matter halo and its neighbouring halos, with velocities measured inside spheres of radius $R$, where $R = 5 \, h^{-1}\mathrm{Mpc}$ in this work.
    \item The \textbf{neighbour statistics}, $\mathcal{D}_5$, is a quantity introduced to probe the statistics of pair separations around each tracer. This estimator is computed as the ratio of the mean separation of halos within spheres of radius $R$ from the main tracer to the population variance. Neighbour statistics is expected to encapsulate information on the small-scale clustering of halos, as they are specifically designed to capture and compress the first two moments of the distribution of pair separations.    
    \item The \textbf{local halo overdensity}, $\Delta_5$, quantifies the density of tracers within a specific region relative to the expected density in a random distribution, based on populations associated with the relative Mach number. $\Delta_5$ is computed by counting the number of tracers used to measure either the relative Mach number or neighbour statistics, and then taking the logarithm of the ratio between the actual number of tracers within a given sphere and the expected number of tracers in a sphere of the same size if they were randomly distributed.
    \item The \textbf{tidal anisotropy}, $\mathcal{T}_A$, measures the degree of anisotropy in the dark matter density field, capturing the directional dependence of gravitational tidal forces. It is characterised by the eigenvalues $\{\lambda_i\}$ of the tidal tensor, which describe how matter is stretched or compressed along different axes due to the large-scale structure of the cosmic web. In this work, $\mathcal{T}_A$ is computed from the invariants of the tidal field, based on the set of eigenvalues derived from the cosmic web classification \citep[see, e.g.,][]{2007MNRAS.375..489H, 2009LNP...665..291V, Balaguera2024}.
\end{itemize}

Figure~\ref{fig:figure5p5} shows the mean scaling relations between halo number density and various halo properties across different MG cosmologies at redshift $z=0.5$. The properties examined include a) $V_{max}$, b) $\mathcal{E}_h$, c) $\mathcal{M}_5$, d) $\Delta_5$, e) $\mathcal{D}_5$ and f) $\mathcal{T}_A$, as labelled in the figure. The error bars represent the uncertainty in the mean for each number density bin. We observe that for $V_{\text{max}}$, all models for the selected samples show a nearly identical relationship with halo number density. In fact, $V_{\text{max}}$ appears to be largely unaffected by modifications of gravity, with any deviations, if present, remaining well below 1\% compared to the $\Lambda$CDM model. However, when considering the ellipticity of halos, we find that only the $F5$ family of models (power indices 1 and 2, represented by the blue and yellow lines, respectively in Fig.~\ref{fig:figure5p5}) shows deviations greater than 1\% below the expectations of $\Lambda$CDM for low-mass halos with densities in the range $\log{\bar{n}} \sim 6 - 8$. This roughly corresponds to halo masses in the range $[1.20 \times 10^{12},~1.62 \times 10^{13}]~h^{-1}\mathrm{M}_\odot$ in the $\Lambda$CDM cosmology.
Below this number density, no significant deviations are observed in any of the models. In fact, all deviations remain consistent within the error bars, as seen in panel b) of Fig.~\ref{fig:figure5p5}, where the shaded regions highlight this consistency. A similar behaviour is observed in the local Mach number (panel c of Fig.~\ref{fig:figure5p5}), where deviations only appear for halos with a number density of $\log{\bar{n}}\geq-8$. In particular, we see a monotonic increase in the ratio of $\mathcal{M}_5$ (see lower panel of c) starting at $\log{\bar{n}}\gtrsim-9$, with departures from $\Lambda$CDM above 1\% for $\log{\bar{n}}>-8$. For the $F6$ family of MG models, no significant deviations are observed across any of the number densities considered. Moving on to the next property analysed, we observe that for the local overdensity, there are no significant deviations in the $F6$ models compared to $\Lambda$CDM; all deviations remain below 1\%. In contrast, the $F5$ family of models shows clear differences relative to $\Lambda$CDM. Specifically, the ratio $\Delta_5^\mathrm{MG} / \Delta_5^{\Lambda\text{CDM}}$ decreases monotonically as the number density increases. The trend is identical for both $F5$ models, indicating no distinction in the impact of the power index of modified gravity. This behaviour of $\Delta_5$ is expected, as stronger modifications to gravity lead to faster expansion of density peaks, resulting in more massive structures at later times, which in turn increases the mean halo number density above a given halo mass threshold. For a comparison of scaling relationships with a different mass definition and threshold in low-resolution $f(R)$ simulations, see \cite{2014MNRAS.440..833A, 2016MNRAS.462.1530A}.

\begin{table}[!htbp]
\begin{center}
\caption{\small{Percent difference of the scaling relations for the non-local halo properties $\mathcal{D}_5$, $\mathcal{M}_5$, and $\Delta_5$ in $f(R)$ models compared to the $\Lambda$CDM model at redshift $z = 0.5$. For each model, the first row corresponds to $\mathcal{D}_5$, the second to $\mathcal{M}_5$, and the third to $\Delta_5$.
}}
\label{table:MGdescription}
\begin{adjustbox}{width=\linewidth,center}
\begin{tabular}{ccccclccccc}
\cline{1-5} \cline{7-11}
$\bar{n}$           &                                   &                                   &                                   &                                   &  & $\bar{n}$           &                                   &                                   &                                   &                                   \\
\multicolumn{1}{l}{{[}$(h^{-1} \mathrm{Mpc})^{-3}${]}} & \multirow{-2}{*}{\textbf{$F5_1$}} & \multirow{-2}{*}{\textbf{$F6_1$}} & \multirow{-2}{*}{\textbf{$F5_2$}} & \multirow{-2}{*}{\textbf{$F6_2$}} &  & \multicolumn{1}{l}{{[}$(h^{-1} \mathrm{Mpc})^{-3}${]}} & \multirow{-2}{*}{\textbf{$F5_1$}} & \multirow{-2}{*}{\textbf{$F6_1$}} & \multirow{-2}{*}{\textbf{$F5_2$}} & \multirow{-2}{*}{\textbf{$F6_2$}} \\ \cline{1-5} \cline{7-11} 
                                                       & \cellcolor[HTML]{C0C0C0}-3.26     & \cellcolor[HTML]{C0C0C0}1.93      & \cellcolor[HTML]{C0C0C0}-0.58     & \cellcolor[HTML]{C0C0C0}-2.98     &  &                                                        & \cellcolor[HTML]{C0C0C0}1.09      & \cellcolor[HTML]{C0C0C0}0.43      & \cellcolor[HTML]{C0C0C0}0.99      & \cellcolor[HTML]{C0C0C0}0.51      \\
-13.15                                                 & \cellcolor[HTML]{EFEFEF}7.19      & \cellcolor[HTML]{EFEFEF}-3.28     & \cellcolor[HTML]{EFEFEF}-1.34     & \cellcolor[HTML]{EFEFEF}-1.71     &  & -9.61                                                  & \cellcolor[HTML]{EFEFEF}-0.70     & \cellcolor[HTML]{EFEFEF}-0.61     & \cellcolor[HTML]{EFEFEF}-1.09     & \cellcolor[HTML]{EFEFEF}-0.23     \\
                                                       & -0.71                             & 3.72                              & 0.49                              & 1.12                              &  &                                                        & -2.88                             & -0.56                             & -2.60                             & -0.23                             \\ \cline{1-5} \cline{7-11} 
                                                       & \cellcolor[HTML]{C0C0C0}-8.86     & \cellcolor[HTML]{C0C0C0}-8.05     & \cellcolor[HTML]{C0C0C0}-7.31     & \cellcolor[HTML]{C0C0C0}-12.94    &  &                                                        & \cellcolor[HTML]{C0C0C0}0.53      & \cellcolor[HTML]{C0C0C0}0.33      & \cellcolor[HTML]{C0C0C0}0.33      & \cellcolor[HTML]{C0C0C0}-0.01     \\
-12.79                                                 & \cellcolor[HTML]{EFEFEF}1.59      & \cellcolor[HTML]{EFEFEF}12.67     & \cellcolor[HTML]{EFEFEF}-5.91     & \cellcolor[HTML]{EFEFEF}0.76      &  & -9.26                                                  & \cellcolor[HTML]{EFEFEF}0.37      & \cellcolor[HTML]{EFEFEF}0.21      & \cellcolor[HTML]{EFEFEF}0.42      & \cellcolor[HTML]{EFEFEF}0.06      \\
                                                       & -2.30                             & -0.59                             & 0.08                              & 4.41                              &  &                                                        & -3.36                             & -1.16                             & -3.05                             & -0.79                             \\ \cline{1-5} \cline{7-11} 
                                                       & \cellcolor[HTML]{C0C0C0}9.64      & \cellcolor[HTML]{C0C0C0}-1.30     & \cellcolor[HTML]{C0C0C0}0.44      & \cellcolor[HTML]{C0C0C0}0.41      &  &                                                        & \cellcolor[HTML]{C0C0C0}0.71      & \cellcolor[HTML]{C0C0C0}0.14      & \cellcolor[HTML]{C0C0C0}0.50      & \cellcolor[HTML]{C0C0C0}0.19      \\
-12.44                                                 & \cellcolor[HTML]{EFEFEF}1.37      & \cellcolor[HTML]{EFEFEF}-3.31     & \cellcolor[HTML]{EFEFEF}9.49      & \cellcolor[HTML]{EFEFEF}0.18      &  & -8.91                                                  & \cellcolor[HTML]{EFEFEF}0.54      & \cellcolor[HTML]{EFEFEF}-0.04     & \cellcolor[HTML]{EFEFEF}0.45      & \cellcolor[HTML]{EFEFEF}0.26      \\
                                                       & -4.77                             & -2.10                             & 0.17                              & -1.76                             &  &                                                        & -3.28                             & -1.04                             & -3.23                             & -0.27                             \\ \cline{1-5} \cline{7-11} 
                                                       & \cellcolor[HTML]{C0C0C0}0.54      & \cellcolor[HTML]{C0C0C0}-1.86     & \cellcolor[HTML]{C0C0C0}-2.59     & \cellcolor[HTML]{C0C0C0}-0.63     &  &                                                        & \cellcolor[HTML]{C0C0C0}1.15      & \cellcolor[HTML]{C0C0C0}0.12      & \cellcolor[HTML]{C0C0C0}0.71      & \cellcolor[HTML]{C0C0C0}0.28      \\
-12.09                                                 & \cellcolor[HTML]{EFEFEF}0.63      & \cellcolor[HTML]{EFEFEF}1.65      & \cellcolor[HTML]{EFEFEF}3.29      & \cellcolor[HTML]{EFEFEF}4.71      &  & -8.55                                                  & \cellcolor[HTML]{EFEFEF}1.02      & \cellcolor[HTML]{EFEFEF}0.17      & \cellcolor[HTML]{EFEFEF}0.68      & \cellcolor[HTML]{EFEFEF}0.09      \\
                                                       & -0.48                             & 1.54                              & -2.01                             & 2.48                              &  &                                                        & -3.00                             & -0.96                             & -3.01                             & -0.03                             \\ \cline{1-5} \cline{7-11} 
                                                       & \cellcolor[HTML]{C0C0C0}-0.57     & \cellcolor[HTML]{C0C0C0}0.10      & \cellcolor[HTML]{C0C0C0}-0.38     & \cellcolor[HTML]{C0C0C0}1.49      &  &                                                        & \cellcolor[HTML]{C0C0C0}1.00      & \cellcolor[HTML]{C0C0C0}-0.09     & \cellcolor[HTML]{C0C0C0}1.11      & \cellcolor[HTML]{C0C0C0}-0.02     \\
-11.73                                                 & \cellcolor[HTML]{EFEFEF}0.39      & \cellcolor[HTML]{EFEFEF}0.11      & \cellcolor[HTML]{EFEFEF}-0.55     & \cellcolor[HTML]{EFEFEF}-0.17     &  & -8.20                                                  & \cellcolor[HTML]{EFEFEF}0.83      & \cellcolor[HTML]{EFEFEF}0.16      & \cellcolor[HTML]{EFEFEF}1.01      & \cellcolor[HTML]{EFEFEF}-0.03     \\
                                                       & -1.13                             & -1.72                             & -2.13                             & -0.47                             &  &                                                        & -3.50                             & -0.91                             & -2.87                             & -0.22                             \\ \cline{1-5} \cline{7-11} 
                                                       & \cellcolor[HTML]{C0C0C0}2.80      & \cellcolor[HTML]{C0C0C0}1.84      & \cellcolor[HTML]{C0C0C0}3.09      & \cellcolor[HTML]{C0C0C0}1.27      &  &                                                        & \cellcolor[HTML]{C0C0C0}1.44      & \cellcolor[HTML]{C0C0C0}0.24      & \cellcolor[HTML]{C0C0C0}1.33      & \cellcolor[HTML]{C0C0C0}-0.14     \\
-11.38                                                 & \cellcolor[HTML]{EFEFEF}-2.97     & \cellcolor[HTML]{EFEFEF}1.40      & \cellcolor[HTML]{EFEFEF}-0.19     & \cellcolor[HTML]{EFEFEF}1.17      &  & -7.85                                                  & \cellcolor[HTML]{EFEFEF}1.21      & \cellcolor[HTML]{EFEFEF}0.23      & \cellcolor[HTML]{EFEFEF}0.60      & \cellcolor[HTML]{EFEFEF}-0.09     \\
                                                       & -2.45                             & -1.25                             & -1.63                             & -0.87                             &  &                                                        & -3.65                             & -0.83                             & -3.43                             & -0.46                             \\ \cline{1-5} \cline{7-11} 
                                                       & \cellcolor[HTML]{C0C0C0}1.20      & \cellcolor[HTML]{C0C0C0}2.89      & \cellcolor[HTML]{C0C0C0}1.63      & \cellcolor[HTML]{C0C0C0}1.02      &  &                                                        & \cellcolor[HTML]{C0C0C0}1.92      & \cellcolor[HTML]{C0C0C0}0.30      & \cellcolor[HTML]{C0C0C0}1.69      & \cellcolor[HTML]{C0C0C0}-0.10     \\
-11.03                                                 & \cellcolor[HTML]{EFEFEF}0.66      & \cellcolor[HTML]{EFEFEF}1.30      & \cellcolor[HTML]{EFEFEF}1.29      & \cellcolor[HTML]{EFEFEF}0.95      &  & -7.49                                                  & \cellcolor[HTML]{EFEFEF}1.22      & \cellcolor[HTML]{EFEFEF}0.12      & \cellcolor[HTML]{EFEFEF}1.01      & \cellcolor[HTML]{EFEFEF}-0.16     \\
                                                       & -1.99                             & 0.63                              & -1.54                             & 0.05                              &  &                                                        & -3.27                             & -0.68                             & -2.93                             & -0.27                             \\ \cline{1-5} \cline{7-11} 
                                                       & \cellcolor[HTML]{C0C0C0}1.42      & \cellcolor[HTML]{C0C0C0}1.06      & \cellcolor[HTML]{C0C0C0}0.58      & \cellcolor[HTML]{C0C0C0}1.46      &  &                                                        & \cellcolor[HTML]{C0C0C0}2.17      & \cellcolor[HTML]{C0C0C0}0.65      & \cellcolor[HTML]{C0C0C0}1.83      & \cellcolor[HTML]{C0C0C0}0.35      \\
-10.67                                                 & \cellcolor[HTML]{EFEFEF}2.12      & \cellcolor[HTML]{EFEFEF}0.86      & \cellcolor[HTML]{EFEFEF}0.98      & \cellcolor[HTML]{EFEFEF}1.14      &  & -7.14                                                  & \cellcolor[HTML]{EFEFEF}1.63      & \cellcolor[HTML]{EFEFEF}0.35      & \cellcolor[HTML]{EFEFEF}1.26      & \cellcolor[HTML]{EFEFEF}0.34      \\
                                                       & -2.19                             & 0.35                              & -1.71                             & 0.81                              &  &                                                        & -3.40                             & -0.56                             & -3.09                             & 0.17                              \\ \cline{1-5} \cline{7-11} 
                                                       & \cellcolor[HTML]{C0C0C0}0.96      & \cellcolor[HTML]{C0C0C0}-0.32     & \cellcolor[HTML]{C0C0C0}-0.05     & \cellcolor[HTML]{C0C0C0}-0.42     &  &                                                        & \cellcolor[HTML]{C0C0C0}1.87      & \cellcolor[HTML]{C0C0C0}0.33      & \cellcolor[HTML]{C0C0C0}1.60      & \cellcolor[HTML]{C0C0C0}-0.13     \\
-10.32                                                 & \cellcolor[HTML]{EFEFEF}0.44      & \cellcolor[HTML]{EFEFEF}-0.10     & \cellcolor[HTML]{EFEFEF}0.74      & \cellcolor[HTML]{EFEFEF}-0.09     &  & -6.79                                                  & \cellcolor[HTML]{EFEFEF}1.39      & \cellcolor[HTML]{EFEFEF}0.27      & \cellcolor[HTML]{EFEFEF}1.05      & \cellcolor[HTML]{EFEFEF}-0.01     \\
                                                       & -2.93                             & -1.11                             & -3.39                             & -0.66                             &  &                                                        & -3.88                             & -1.01                             & -3.47                             & -0.40                             \\ \cline{1-5} \cline{7-11} 
                                                       & \cellcolor[HTML]{C0C0C0}0.48      & \cellcolor[HTML]{C0C0C0}1.04      & \cellcolor[HTML]{C0C0C0}-0.11     & \cellcolor[HTML]{C0C0C0}-0.14     &  &                                                        & \cellcolor[HTML]{C0C0C0}2.04      & \cellcolor[HTML]{C0C0C0}0.43      & \cellcolor[HTML]{C0C0C0}1.79      & \cellcolor[HTML]{C0C0C0}0.09      \\
-9.97                                                  & \cellcolor[HTML]{EFEFEF}-0.27     & \cellcolor[HTML]{EFEFEF}0.24      & \cellcolor[HTML]{EFEFEF}0.18      & \cellcolor[HTML]{EFEFEF}0.29      &  & -6.43                                                  & \cellcolor[HTML]{EFEFEF}1.60      & \cellcolor[HTML]{EFEFEF}0.27      & \cellcolor[HTML]{EFEFEF}1.22      & \cellcolor[HTML]{EFEFEF}0.11      \\
                                                       & -2.97                             & 0.24                              & -1.78                             & 0.20                              &  &                                                        & -3.79                             & -0.83                             & -3.28                             & -0.22                             \\ \cline{1-5} \cline{7-11} 
\end{tabular}
\end{adjustbox}
\end{center}
\end{table}

\begin{figure*}
  \centering
  \includegraphics[width=\linewidth]{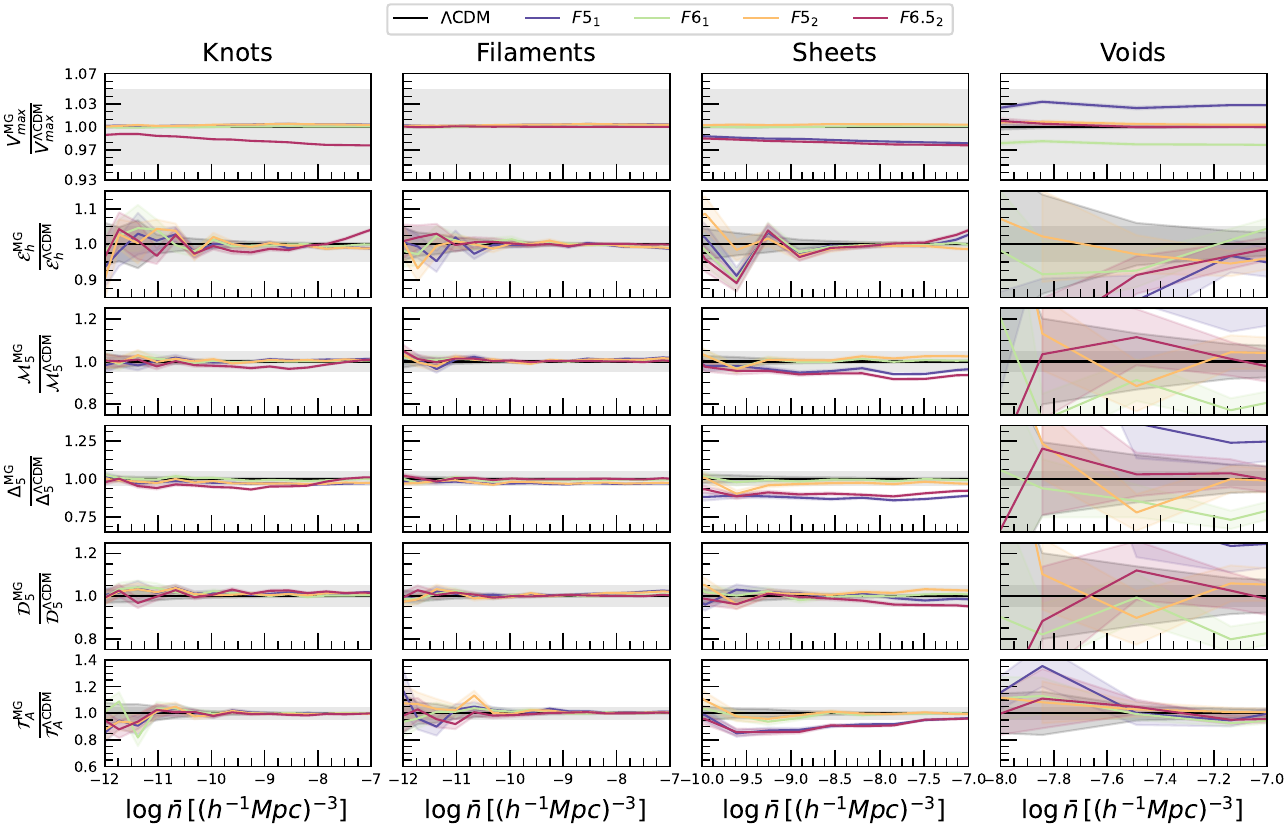}
     \caption{
     \small{Relative mean scaling relations in the cosmic web. The lines represent different halo properties in $f(R)$ cosmologies as a function of number density at redshift $z = 0.5$. The rows show the ratio $\Theta^{\mathrm{MG}} / \Theta^{\Lambda\mathrm{CDM}}$ for the properties $V_{\mathrm{max}}$, $\mathcal{E}_h$, $\mathcal{M}_5$, $\Delta_5$, $\mathcal{D}_5$, and $\mathcal{T}_A$. Each column corresponds to a different element of the cosmic web classification, i.e., knots, filaments, sheets and voids. The error bars represent the uncertainty in the mean for each number density bin.}
     }
        \label{fig:figure7}
\end{figure*}

For the neighbour statistics, $\mathcal{D}_5$, we observe that the MG models generally follow the overall trend of the $\Lambda$CDM model. In the lower panel of Fig.~\ref{fig:figure5p5}e, the ratio of neighbour statistics of MG to $\Lambda$CDM, $\mathcal{D}_5^{\text{MG}} / \mathcal{D}_5^{\text{CDM}}$, fluctuates around 1\%, indicating only minor deviations from the reference model. The $F6$ family of models remains virtually indistinguishable from $\Lambda$CDM, whereas the neighbour statistics in $F5$ models exhibit some significant variations. In this case, $\mathcal{D}_5$ shows fluctuations at low number densities ($\log \bar{n} < -12$), with deviations reaching up to approximately 5\% relative to the $\Lambda$CDM. As the number density increases, the ratio stabilizes, remaining within 1\%, which suggests that the $F5$ models align closely with standard gravity at higher densities. A slight turnaround is observed around $\log \bar{n} \approx -9.5$, where the ratio dips just below 1\% before recovering. Beyond this point, the neighbour statistics consistently remain below 3\%, further indicating their close alignment with $\Lambda$CDM at higher densities. Lastly, and equally important, we analyse the triaxiality of MG halos. In the Fig.~\ref{fig:figure5p5}f, we observe that the results from MG models align with the overall trend of halo triaxiality of standard gravity. Focusing on the ratio of triaxiality (see the lower panel), at low number densities ($\log \bar{n} < -10$), the measurements are scattered enough to not show any conclusive departures from $\Lambda$CDM. However, beyond this threshold, a small deviation becomes apparent in all MG models, remaining constant up to $\log \bar{n} = -6$. This deviation is around 1\% for the $F5_1$ model and below 1\% for the $F5_2$ and the remaining $F6$ models. Notably, triaxiality is the only property where we observe an effect of the power index of MG.

Table~\ref{table:MGdescription} presents the percent differences of the scaling relations for three non-local halo properties in $f(R)$ models relative to the $\Lambda$CDM model, i.e., $100(\Theta^{\mathrm{MG}} / \Theta^{\Lambda\mathrm{CDM}}-1)$[\%], as a function of the mean number density. The table includes the properties $\mathcal{D}_5$ (first row, dark grey), $\mathcal{M}_5$ (second row, medium grey), and $\Delta_5$ (third row, white), which are the ones that exhibit the largest deviations from standard gravity across different number density. Notably, these deviations are most pronounced for low-mass halos (higher number densities). We see that the quantity showing the largest deviation from $\Lambda$CDM is the local overdensity across most of the number density range, as already shown in Fig.~\ref{fig:figure5p5}. Specifically, for model $F5_1$, $\Delta_5$ is approximately 4\% lower than the value expected in $\Lambda$CDM at $\bar{n} \approx 6.43 \, [h^{-1} \, \mathrm{Mpc}]^{-3}$, while considering the power index $n=2$, this difference decreases to about 3.3\%. For the $F6$ models, the local overdensities remain below 1\% over a wide range of number densities.

Figure~\ref{fig:figure7} shows the relative mean scaling relations in the cosmic web for $f(R)$ cosmologies as a function of number density at redshift $z = 0.5$. Each row displays the behaviour of a property —$V_{\mathrm{max}}$, $\mathcal{E}_h$, $\mathcal{M}_5$, $\Delta_5$, $\mathcal{D}_5$, and $\mathcal{T}_A$— across different cosmic web environments: knots, filaments, sheets, and voids, as labelled in the columns \footnote{As in \cite{Balaguera2024}, this classification is based on the eigenvalues of the tidal field setting a null threshold.}. As in previous figures, the error bars represent the uncertainty in the mean for each number density bin, and the gray region indicate a 5\% error reference. We observe that sheets and voids are the most sensitive cosmic environments, showing notable differences when modifications to gravity are considered. The other environments, knots and filaments, are relatively insensitive to variations in the gravity model, with any discrepancies from $\Lambda$CDM remaining below 5\% for all the $f(R)$ models considered. In the case of sheets, however, the differences are above 5\% compared to $\Lambda$CDM, especially for properties like the local Mach number, local overdensity, neighbour statistics, and tidal anisotropy of halos. This can be attributed to the fact that sheets are regions of intermediate density, positioned between dense clusters and less dense voids. Consequently, the impact of modified gravity on the local overdensity causes halos in sheets to evolve differently compared to those in more extreme environments. In sheets, the halo abundance is neither as high as in clusters nor as sparse as in voids, indicating a balanced density of halos.

An interesting feature observed in the properties of halos within sheets, such as tidal anisotropy, local overdensity, and Mach number, is that this environment appears to maximize the impact of the power index of MG as shown in Fig.~\ref{fig:figure7}. For model $F5_2$, these properties closely align with $\Lambda$CDM, with deviations well below 5\%. However, for $F5_1$, significant differences emerge with respect to $\Lambda$CDM: 5\% in $\mathcal{M}_5$ at $\log\bar{n} \sim -7.5$; 12\% in $\Delta_5$ at $\log\bar{n} \sim -8$; and 15\% in $\mathcal{T}_A$ at $\log\bar{n} \sim -9.5$. A similar trend is observed when comparing $F6_1$ with $F6_2$, where differences with respect to $\Lambda$CDM are 7\% in $\mathcal{M}_5$; 10\% in $\Delta_5$, and 15\% in $\mathcal{T}_A$, for the same number density values as those for $F5$ models. These findings suggest a potential connection between the power index of $f(R)$ and the evolution of large-scale structures in sheets. However, a more detailed investigation of this trend is beyond the scope of this paper and may be the subject of future work.

\begin{figure}
  \centering
  \includegraphics[width=\linewidth]{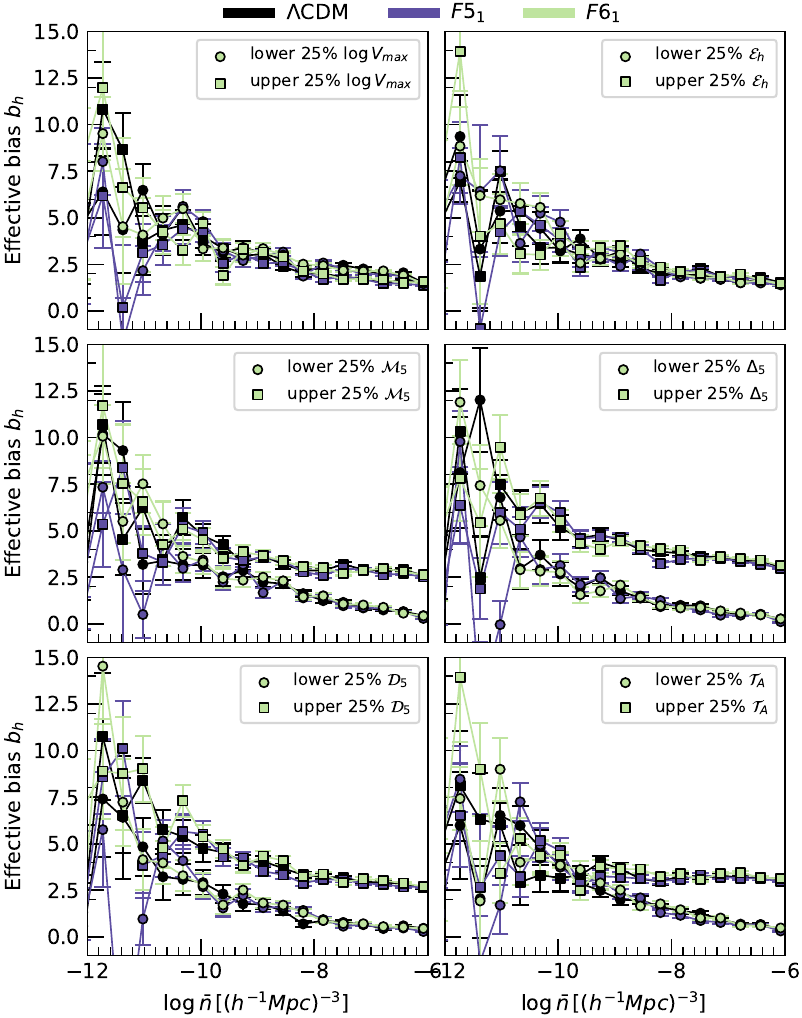}
     \caption{
     \small{Secondary halo bias: mean halo effective bias $\langle b_h|\bar{n}\rangle_q^{\theta}$ measured in bins of the halo (log) number density using a number of halo properties as secondary properties $\theta$, namely, $\log V_{\text{max}}$, $\mathcal{E}_h$, $\mathcal{M}_5$, $\Delta_5$, $\mathcal{D}_5$, $\mathcal{T}_A$. In each number density-bin, the sample has been divided in quartiles $q$ of the property $\theta$. We show results from the lower (first) quartile (circles) and the upper (third) quartile (squares), for two modified gravity models $F5_1$ (light blue) and $F6_1$ (light green) alongside results from the $\Lambda$CDM reference (black), all of them at the same redshift $z=0.5$.}
     }
    \label{fig:figure2p1}
\end{figure}
\begin{figure}
  \centering
  \includegraphics[width=\linewidth]{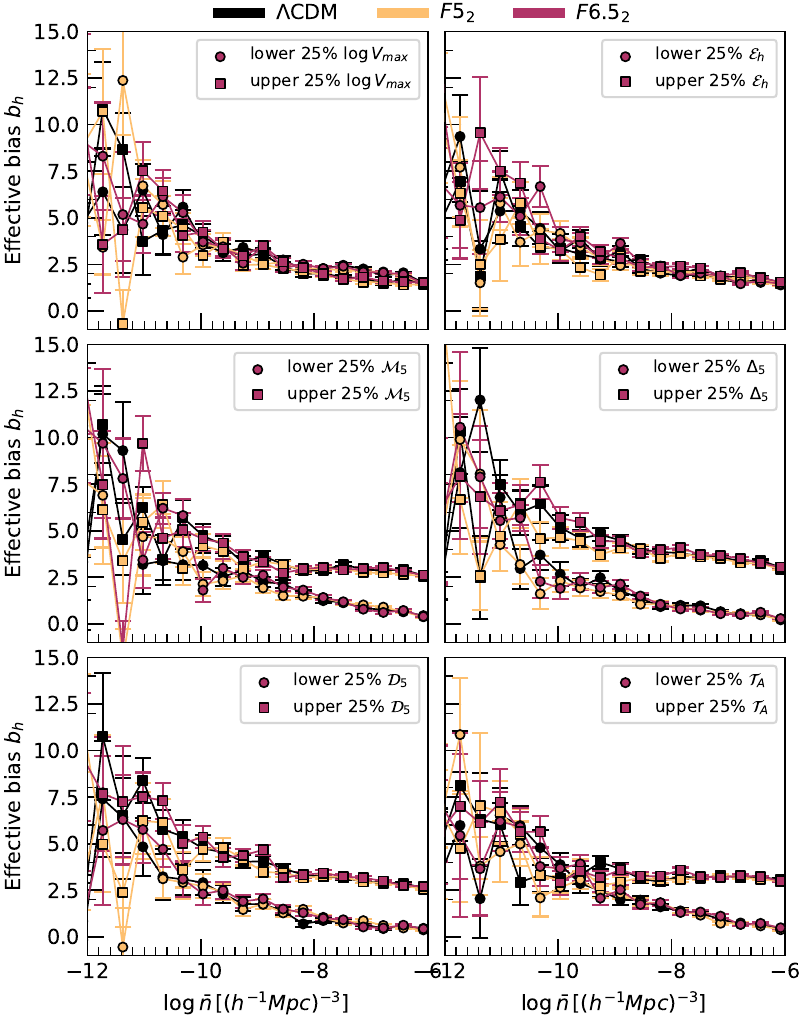}
     \caption{
     \small{Same as Fig.~\ref{fig:figure2p1} but for the remaining MG models, $F5_2$ (orange) and $F6.5_2$ (dark purple) alongside results from the $\Lambda$CDM reference (black), all of them at the same redshift $z=0.5$.}
     }
    \label{fig:figure2p2}
\end{figure}
\begin{figure*}
  \centering
  \includegraphics[width=\linewidth]{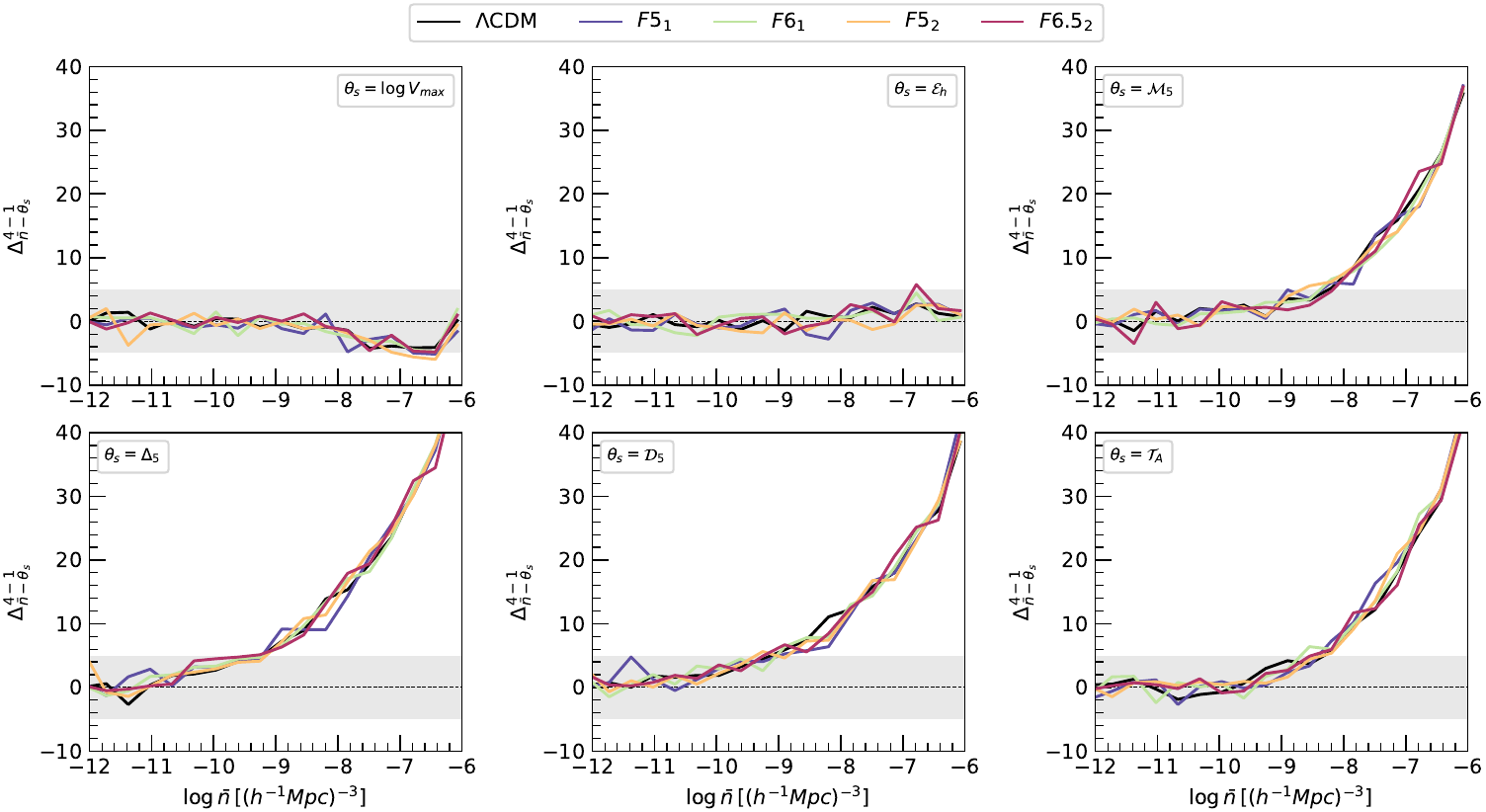}
     \caption{
     \small{Significance of the secondary halo bias as a function of number density, computed using Eq.~\eqref{eq:significance} for different $f(R)$ models and for the secondary properties: $V_{\mathrm{max}}$, $\mathcal{E}_h$, $\mathcal{M}_5$, $\Delta_5$, $\mathcal{D}_5$, and $\mathcal{T}_A$. The dark-shaded stripe denotes a strip of $\Delta^{4-1}_{p-s} \pm 5$ and the horizontal line marks $\Delta^{4-1}_{p-s} = 0$.}
     }
    \label{fig:figure2p3}
\end{figure*}

\subsection{Secondary bias}
The signal of the clustering of dark matter halos has been shown to depend on various secondary halo properties, rather than only halo mass. The secondary bias refers to the clustering of a population of matter tracers selected based on a secondary property, as a function of a primary property—in this case, the halo number density. The secondary bias is calculated by dividing the sample into quartiles of the secondary property and measuring the mean effective halo bias in bins of halo number density \citep[for details on the procedure, see e.g.,][]{MonteroDorta2020B, MonteroRodriguez2024,Balaguera2024}. In this work, we focus on six representative secondary properties: $V_{\text{max}}$, $\mathcal{E}_h$, $M_5$, $\Delta_5$, $D_5$, and $\mathcal{T}_a$.

Figures~\ref{fig:figure2p1} and \ref{fig:figure2p2} show the measurements of mean effective bias in the first (lower) and fourth (upper) quartiles of all these properties for $f(R)$ models with power index $n=1$ and $n=2$, respectively, with the $\Lambda$CDM results as reference in both cases (black data points). All the properties shown in these figures exhibit signatures of secondary bias, with significance that vary with number density but overall remains stable across the different $f(R)$ models. In fact, we see that the $f(R)$ models considered, closely mimic the secondary bias of $\Lambda$CDM. This is evidenced by the agreement in bias signature between the two quartiles, indicating that these type of halo properties do not exhibit significant secondary bias, e.g., $V_{max}$ and $\mathcal{E}_h$. On the other hand, properties such as $\mathcal{M}_5$, $\Delta_5$, $\mathcal{D}_5$, and $\mathcal{T}_A$ display, as expected, clear secondary bias signatures \citep[see, for example][for a comparison]{Balaguera2024,Balaguera2024_2}.

To quantify the extent of deviation in the secondary bias of $f(R)$ models with respect to $\Lambda$CDM, we compute the significance of the signal, defined as
\begin{equation}\label{eq:significance}
\Delta_{p-s}^{i-j} \equiv \frac{\left\langle b_h \mid \theta_p\right\rangle_i^{(s)}-\left\langle b_h \mid \theta_p\right\rangle_j^{(s)}}{\sqrt{\sigma_{s, i}^2+\sigma_{s, j}^2}},
\end{equation}
where $\langle b_h | \theta_p \rangle_i^{(s)}$ represents the mean sample bias measured in bins of the primary property $\theta_p$ for the $i$-th quartile of the secondary property $\theta_s$. The standard error of the mean bias in the $i$-th quartile of the secondary property is denoted by $\sigma_{s,i}$. In our analysis, $i=4$ and $j=1$ correspond to the fourth (upper) and first (lower) quartiles, respectively. Therefore, the significance is $\Delta^{4-1}_{p-s}$, with $p$ representing $\bar{n}$ and $s$ being one of the secondary properties: $V_{\mathrm{max}}$, $\mathcal{E}_h$, $\mathcal{M}_5$, $\Delta_5$, $\mathcal{D}_5$, and $\mathcal{T}_A$. One statistical indicator of a secondary halo bias signature is a significance value considerably higher than 1. This is important because some secondary qualities may have a large signal within a specific mass range (number density), but this signal might not be statistically significant. Figure~\ref{fig:figure2p3} shows the statistical significance of the secondary halo bias as a function of halo number density for the different $f(R)$ models (coloured lines) and $\Lambda$CDM (black lines).

Properties such as $\mathcal{M}_5$, $\Delta_5$, $\mathcal{D}_5$, and $\mathcal{T}_A$ clearly show a secondary bias, consistent with the results from \cite{Balaguera2024} for $\Lambda$CDM using the UNIT simulation. Figure~\ref{fig:figure2p3} extends these findings to a variety of $f(R)$ gravity models, where the strength of the secondary bias signal remains similar across all the models considered. The figure illustrates how the statistical significance changes with halo number density for the selected properties, with an uncertainty range of $\Delta_{p-s}^{4-1} \pm 5$ included to guide the reader. Notably, as number density increases, the significance of the bias increases monotonically. Figure~\ref{fig:figure2p3} shows that distinguishing the $f(R)$ models from $\Lambda$CDM require measurements with (sub)percent precision in multiple number density bins, especially those for very massive and rare halos. Fortunately, near-future surveys expect to achieve this.

Our analysis reveals that, across all $f(R)$ models studied, $V_{\mathrm{max}}$ and $\mathcal{E}_h$ do not exhibit secondary bias, as expected. The remaining properties, instead, display a similar statistical significance of secondary bias. At a reference number density of $\sim10^{-7}$ —consistent with the expected comoving number densities of surveys such as DESI and Euclid \citep[see e.g.,][]{2024arXiv240513491E, 2023AJ....165...58Z}— we do observe notable deviations. Specifically, the significance levels are 18 units for $\mathcal{M}_5$, 28 for $\Delta_5$, 20 for $\mathcal{D}_5$, and 22 for $\mathcal{T}_A$, consistently across all $f(R)$ models. These results hold irrespective of the power index or the $f_{R0}$ parameter (see Fig.~\ref{fig:figure2p3}).

To quantify the bias, we employ the methodology described by \cite{2018MNRAS.476.3631P}, which was later implemented by \cite{Contreras2021a} and \cite{Balaguera2024}. This approach is based on an object-by-object estimator to calculate the halo bias. We apply this method to assess the signature of the secondary bias, utilizing dark matter halos and their properties obtained with the \texttt{ROCKSTAR} halo finder. Our results extend the findings reported in \cite{Balaguera2024} on secondary halo bias to the context of $f(R)$ cosmologies for the first time, while also validating the method in the $\Lambda$CDM scenario.

 \section{Conclusions}\label{sec:conclusions}
In this work, we present the measurements of primary and secondary bias for dark matter halos, extracted from the high-resolution simulations of MG cosmologies introduced by \cite{2024A&A...690A..27G}. The MG models correspond to the Hu-Sawicki parametrization of the $f(R)$ function. We consider two sets of models, each defined by specific combinations of the $f_{R0}$ and $n$ parameters, chosen to represent models that remain consistent with current observational constraints. Specifically, we investigate four MG models in addition to the reference $\Lambda$CDM model, $(|f_{R0}|,\,n) \in \{(10^{-5},\,1),\,(10^{-6},\,1),\,(10^{-5},\,2),\,(10^{-6.5},\,2)\}$, which are denoted as $\{F5_1, F6_1, F5_2, F6.5_2\}$ throughout this work.

We have explored the signal of bias using a number of halo properties derived from the halo finder. Among these, intrinsic properties are the maximum circular velocity and halo ellipticity. As a primary property, we have used the mean number density (interpolated from each $f(R)$ catalog) instead of halo mass, as the latter depends also on the MG models used. We have also analysed the impact on environmental properties such as the Mach number and the neighbour statistics, introduced by \cite{Balaguera2024}. 
Our main conclusions can be summarised as follows:

\begin{itemize}
    \item The relative difference in the effective bias among the $f(R)$ models when compared to $\Lambda$CDM, encodes information of the power index impact on the bias. For a fixed power index $n=1$, the halo bias decreases as $|f_{R0}|$ increases, with maximum deviations of about 5\%. This occurs because halos with a fixed number density become more abundant, leading to a bias reduction. This result is consistent with previous studies, e.g., \citep{2009PhRvD..79h3518S}, where the largest deviations from GR appear in low-mass halos. However, when the power index is fixed at $n = 2$, the suppression of the halo bias with respect to $\Lambda$CDM is no longer evident. Instead, the bias remains consistent across models with different $|f_{R0}|$, such as $F5_2$ and $F6.5_2$, except for rare halos with low number densities. The impact of $n$ manifest better in the $F5$ models, where increasing the power index reduces the suppression of the bias across a wide range of number densities.
    \item The scaling relations for the full $f(R)$ catalogs with the properties, $V_{max}$, $\mathcal{E}_h$, $\mathcal{M}_5$, $\Delta_5$, $\mathcal{D}_5$ and $\mathcal{T}_A$ are presented in Fig.~\ref{fig:figure5p5}. Our analysis reveals that $ V_{\text{max}}$ exhibits a nearly universal relationship with halo number density across all models, showing negligible deviations (<1\%) from $\Lambda$CDM. In contrast, halo ellipticity displays significant deviations (>1\%) only in the $F5$ models (for both power indices) within the low-mass range $[1.20 \times 10^{12},~1.62 \times 10^{13}]~h^{-1}\mathrm{M}_\odot$, while no notable differences arise at lower number densities ($\log{\bar{n}} \sim 6-8$). For the Mach number, departures from $\Lambda$CDM emerge at $\log{\bar{n}} \geq -8$, with a monotonic increase in $\mathcal{M}_5$ ratios beginning at $\log{\bar{n}} \gtrsim -9$, whereas the $F6$ models remain consistent across all densities. Similarly, the local overdensity shows no significant deviations in most models, except for the $F5$ family, where $\Delta_5^\mathrm{MG} / \Delta_5^{\Lambda\text{CDM}}$ decreases monotonically with increasing number density, unaffected by the power index. 
    \item The halo properties studied through scaling relations were analysed across different environments of the cosmic web to quantify their environmental impact. The results show that sheets and voids are the most responsive environments to $f(R)$, exhibiting deviations across multiple halo properties. In these environments, the tidal anisotropy, local overdensity, and Mach number appear to be sensitive to the power index of $f(R)$, with differences exceeding 5\% compared to $\Lambda$CDM, reaching up to 12\% in $\Delta_5$ and 15\% in $\mathcal{T}_A$ for the $F5_1$ model. On the contrary, knots and filaments remain largely unaffected, with any discrepancy below 5\%.     
    We found that, while $F5_2$ closely aligns with $\Lambda$CDM (deviations <5\%), $F5_1$ and $F6_1$ exhibit stronger departures (e.g., 7–15\% in $\mathcal{M}_5$, $\Delta_5$, and $\mathcal{T}_A$), which hints at the impact of the power index on cosmic tracers that inhabit underdense regions. These findings suggest that traces of $f(R)$ gravity are environment-dependent, with sheets and voids potentially serving as probes for deviations from GR.
    \item The measurements of the secondary bias across halo properties (including $V_{\text{max}}$, $E_h$, $\mathcal{M}_5$, $\Delta_5$, $\mathcal{D}_5$, and $\mathcal{T}_A$), show a consistent behaviour that persists across all $f(R)$ models. We found that the $f(R)$ models closely reproduce the secondary bias of $\Lambda$CDM with high fidelity in all cases. We report that while properties such as $V_{\text{max}}$ and halo $\mathcal{E}_h$ have a negligible secondary bias, as expected, $\mathcal{M}_5$, $\Delta_5$, $\mathcal{D}_5$, and $\mathcal{T}_A$ exhibit statistically significant trends, albeit remaining fully compatible with $\Lambda$CDM.
    \item We quantify the extent of the deviation in the secondary bias of $f(R)$ halos relative to $\Lambda$CDM by computing the significance of the signal, accounting for the intrinsic scatter in the measurements. For halo properties such as $\mathcal{M}_5$, $\Delta_5$, $\mathcal{D}_5$, and $\mathcal{T}_A$, we find a statistical significance of the secondary bias consistent with the $\Lambda$CDM results reported by \cite{Balaguera2024}. When extending this analysis to $f(R)$ halos, the signal strength remains remarkably similar across all investigated parameter combinations, regardless of variations in both $f_{R0}$ and the power index. In general, we do not find any statistically significant deviation from $\Lambda$CDM in the secondary halo bias for the different $f(R)$ models explored in this work.
\end{itemize}

We anticipate several extensions of the analysis presented in this work. In particular, the set of simulations we have employed can be combined with various filament and void identification schemes to explore in more detail the sensitivity of the relationship between large-scale bias and the cosmic web to variations in the underlying cosmology. 

\begin{acknowledgements}
J.E. García-Farieta research was financially supported by the project ``Plan Complementario de I+D+i en el área de Astrofísica'' funded by the European Union within the framework of the Recovery, Transformation and Resilience Plan - NextGenerationEU and by the Regional Government of Andalucia (Reference AST22\_00001). ABA acknowledges the Spanish Ministry of Economy
and Competitiveness (MINECO) under the Severo Ochoa program SEV-2015-
0548 grants. A.D. Montero-Dorta thanks Fondecyt for financial support through the Fondecyt Regular 2021 grant 1210612.
\end{acknowledgements}

\bibliographystyle{aa}
\bibliography{refs}
\end{document}